\documentstyle[aps,prl,twocolumn,tighten]{revtex}

\begin{document}

\title{Transmission and Scattering of a Lorentz Gas on a Slab}

\author{Hern\'an Larralde $^{\star \S}$, Fran\c{c}ois Leyvraz $^{\star \S}$, \\ 
Gustavo Mart\'{\i}nez-Mekler $^{\star \S}$, 
Ra\'ul Rechtman $^{\triangle \S}$\cite{rech}, 
Stefano Ruffo $^{\diamond\S}$\cite{ruffo}}

\address{
$\star$ Laboratorio de Cuernavaca, Instituto de F\'{\i}sica, UNAM,
Apdo. Postal 48-3, \\
62251 Cuernavaca, Morelos, M\'exico \\
$\triangle$ Centro de Investigac\'{\i}on en Energ\'{\i}a, UNAM, \\
62580 Temixco, Morelos, M\'exico \\
$\diamond$ Dipartimento di Energetica ``S. Stecco'', Universit\`a di Firenze,
via S. Marta 3, \\
50139 Firenze, Italy and INFN Firenze \\
$\S$ Centro Internacional de Ciencias, Cuernavaca, Morelos, M\'exico
}

\maketitle

\begin{abstract}
We perform numerical scattering experiments on a Lorentz array of
disks centered on a triangular lattice with $L$ columns and study its
transmission and reflection properties. In the finite horizon case,
the motion of the particles may be modeled as simple one dimensional
random walks with absorbing walls for which the scaling of the
transmission and reflection coefficients are known, and agree with
those found numerically. In the infinite horizon case the analogy with
a simple diffusive process is no longer valid. In this case we compare
our results to those expected for a one dimensional L\'evy walk, again
with absorbing walls, for which logarithmic corrections to the scaling
relations appear. These corrections are consistent with the numerical
results. The scaling with $L$ and the symmetry properties of the
forward $\sigma_T (\phi)$ and backward $\sigma_R (\phi)$ differential
cross sections are also studied, and some of their salient features
are discussed.


\end{abstract}

\vspace{0.3cm}
\noindent PACS number(s): 05.40 +j,05.45 +b,05.60 +w

\section{Introduction}
The Lorentz gas is an ensemble of noninteracting point particles which
move freely with elastic reflections from fixed scatterers~\cite{lor05}.
It is a basic model for linearized kinetic equations~\cite{kin} and
its ergodic properties are well known~\cite{ergo}.

In this paper we present the results of numerical experiments in which
a large number of particles are incident on an array of disks centered
on a triangular lattice. The particles are launched initially either
in the $+x$ direction or isotropically towards the array, and are
reflected elastically from the scatterers. The array of disks is
finite in the $x$ direction and infinite in the $y$ direction so we
speak of a ``slab'' of scatterers.  Our slab is characterized by two
parameters, the width of the slab, that is, the number $L$ of columns
of scatterers, and the separation $w$ between them (the disk radius is
unity). The quantities that are measured are the transmission $T$ and
reflection $R$ coefficients, the mean survival time $\tau$ of
particles in the slab and the transmitted $\sigma_T$ and reflected
$\sigma_R$ differential cross sections. These quantities are analyzed
as functions of the parameters characterizing the slab.

If the separation between scatterers $w$ in a triangular lattice is
small, $w<w_c=0.3094\dots$, the length of free motion of the point
particles is bounded, that is, the particles ``see'' a finite
horizon. On the other hand, when $w_c<w$ the length of free motion may
be unbounded, the particles ``see'' an infinite horizon. In this work
we study the scattering properties in both situations, and analyze our
results in terms of the characteristic motion of the particles in each
case. In the finite horizon case, the motion of the particles is known
to be diffusive~\cite{bun81} and the diffusion coefficient can be
estimated with ergodic arguments~\cite{mac83}. Velocity correlation
functions have been shown to decay exponentially~\cite{bun81}, as
confirmed by numerical experiments~\cite{gar94}.  In contrast, for the
infinite horizon case the diffusion coefficient diverges
logarithmically~\cite{bou85,ble92} and correlation functions have a
power law decay~\cite{corr}. We find that there are also fundamental
differences in the scattering properties for each case.

Lorentz gases in finite size geometries have also been introduced
elsewhere with the aim of studying escape rates and their relation to
transport coefficients and fractal repellers~\cite{gas95}. An account
of these results, together with a formulation of the problem in terms
of flux boundary conditions can be found in Ref.~\cite{gas97}.

In Section~\ref{experiment} we introduce all the definitions and
discuss in detail the numerical experiments for the case of a finite
horizon.  Section~\ref{wald} draws an analogy between the Lorentz
scattering experiments in the finite horizon case and the behavior of
one-dimensional diffusive motion with absorbing boundaries. This
serves as a basis for the explanation of the observed scaling laws for
transmission coefficients and survival times.  In
Section~\ref{infinite} we present the results for the infinite horizon
case when the particles are launched initially in the $x$ direction.
These show logarithmic corrections to the scaling laws, which are
consistent with considering the motion of the particles within the
slab as a L\'evy walk.  In Section~\ref{angular} we discuss angular
dependences and symmetries of the transmission and reflection
differential cross sections. In Section~\ref{isotropic} we discuss how
some of the scattering properties are affected by sending
the particles isotropically, i.e. with an incidence angle uniformly
distributed between $-\pi/2$ and $ \pi/2$.  A modified random walk model
displaying these same differences is also briefly discussed. Finally,
Section~\ref{conclusions} is devoted to discussion of the results and
conclusions.

\section{Scattering with a finite horizon}
\label{experiment}

\begin{figure*}
\vspace{5cm}
\includegraphics{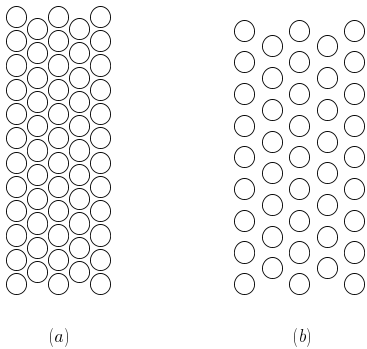}
\caption{\label{fslab} Slabs of scatterers in a triangular array.
(a) Finite horizon, $w=0.3$, with $L=5$. (b) Infinite horizon,
$w=1.$, with $L=5$.}
\end{figure*}

The scatterers we consider are disks of unitary radius centered on a
triangular lattice as shown in Fig.~\ref{fslab}. The distance between
neighboring centers is $2+w$ and the centers lie along $L$ lines
parallel to the $y$ axis.  The slab is finite in the $x$ direction
and infinite in the $y$ direction. A large number $N$ of particles are
incident from the left parallel to the $x$ axis with unit speed.
The particles move freely
except for elastic collisions at the boundary of the disks.  
In the experiments, the dynamics is solved by
considering the motion in the elementary Wigner-Seitz hexagonal cell
where opposite sides are identified. 
Each incident particle has a different impact parameter $b$, defined here
as the distance between the initial position and the horizontal line passing
through the center of the scatterer in the Wigner-Seitz cell. Due to the
symmetry of the slab, it is sufficient to consider b between 0 
and $1+w/2$. The trajectory in the slab is
obtained by unfolding the orbit in the Wigner-Seitz cell.
The cases where particles are incident with
an angle different from zero and the effect of isotropic incidence will be
briefly discussed in Section~\ref{isotropic}. 

In the finite horizon case, $0<w<w_c=(4/\sqrt{3}-2)=0.3094\dots$, 
the particles that enter the
scatterer cannot travel long distances without suffering collisions.
In the infinite horizon case, $w_c<w$, 
the particles may travel arbitrarily far between collisions, 
due to the opening of infinite corridors. If $w_c<w<2$ the infinite corridors
lie at angles of $\pi/6$, $\pi/3$, and $\pi/2$. We will be mainly considering
particles incident parallel to the $x$ axis that cannot 
cross the slab without
collisions when $w<2$.

Since the slab is infinite in the $y$ direction and the collisions are
elastic, every particle that enters the slab must eventually exit it,
except for a set of measure zero which goes asymptotically to periodic
orbits inside the slab (we disregard all zero measure sets
herafter). Thus, in practice, a particle that enters the slab collides
with some of the obstacles and will be ultimately transmitted or
reflected, leaving the slab with an angle $\phi$ measured with respect
to the $+x$ direction. Particles are transmitted if they exit the slab
from the right ($|\phi| < \pi/2$), and are reflected if they exit from
the left ($|\phi| > \pi/2$).

A first characterization of the system is through the computation of
the transmision $T$, and reflection $R$, coefficients. The former is
defined as the fraction of particles that pass through the slab and
exit on the right, and the latter as the particle fraction that exits
the slab on the left (obviously $T+R=1$).  A finer quantity is the
differential scattering cross section $\sigma$ defined by saying that
$\sigma(\phi)d\phi$ is the fraction of particles scattered between
$\phi$ and $\phi + d\phi$. We can separate this quantity in the
transmitted $\sigma_T$ and reflected $\sigma_R$ differential
scattering cross sections by considering $|\phi|<\pi/2$ and
$|\phi|>\pi/2$ respectively.

\begin{figure*}
\vspace{6cm}
\includegraphics{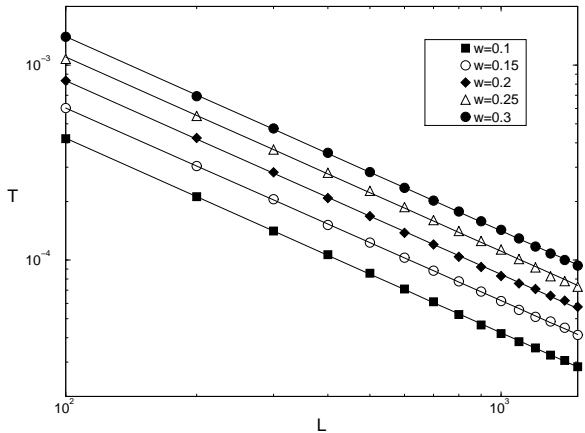}
\caption{\label{ftl} Logarithmic plot of the transmission
coefficient T as a function of the size of the system L, for
different values of $w$. The number N of incident particles
were, for $w=0.1, 4\times 10^7$; for $w=0.15, 2 \times 10^7$, and
for $w=0.2, 0.25, 0.3, 10^7$.  The full lines are the least square
fits, all of them compatible with the theoretical prediction $T \sim
L^{-1}$ to within $1\%$.}
\end{figure*}

In Fig.~\ref{ftl} we show the dependence of the transmission
coefficient $T$ on $L$ for the finite horizon case. These results are
consistent with the scaling law $T \sim L^{-\beta}$ where
$\beta\approx 1$ independently of $w$. This
behavior will be justified in the next section drawing from an analogy
with random walks. Since $R=1-T$ and $T$ scales to zero with $L$, $R$
does not depend on $L$ for $L\gg 1$.  Another quantity of interest is
the mean survival time inside the slab $\tau$. This average time is
evaluated over all the $N$ incident particles no matter if they are
transmitted or reflected. We find that $\tau\sim L^\gamma$ with
$\gamma\approx 1$ as we show in Fig.~\ref{ftau}.

\section{Diffusive behavior in finite size systems}
\label{wald}

In the finite horizon case, rigorous results, convincing evidence and
plausible arguments have been set forth indicating that the motion of
the particles in the Lorentz system can be accurately modeled as a
simple random walk~\cite{bun81,mac83}. In its simplest version the
particles can be viewed as hopping between adjacent ``cages'' in an
essentially uncorrelated fashion, and staying in each cage for a well
defined average time.  For the case under study in this paper it is
not even necessary to consider the random walk process as occuring on
a two dimensional lattice since the quantities we are interested in
can be calculated from the projection of the walk onto the finite
direction $x$.

While the discrete one dimensional random walk on a finite lattice can
be described completely~\cite{fel,weiss}, such a detailed comparison
between the two systems cannot hold. As the random walk is an analogy
to the Lorentz system, the best we can realistically expect to
determine from it is the scaling behavior of the quantities under
study. With this in mind, we choose to evaluate the transmission and
reflection coefficients for the simple random walk problem via the
diffusion equation, with a numerically determined phenomenological
diffusion constant $D$. This equation describes the evolution of the
coarse grained particle density in the system and is expected to be
valid when the system size is much greater than the root mean square
(rms) step length. Once again, since the slab is translationally
invariant along the $y$ axis, the coarse grained particle density
obeys a diffusion equation along $x$.

\begin{figure*}
\vspace{6cm}
\includegraphics{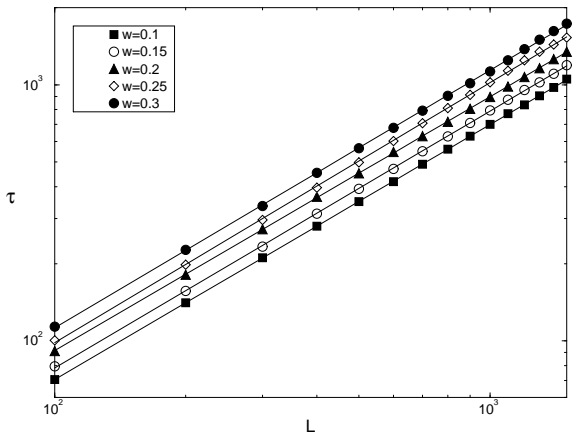}
\caption{ \label{ftau} Logarithmic plot of the survival time $\tau$ vs. $L$
for the same values of $w$ and $N$ as in Fig.~\protect{\ref{ftl}}.
The full lines are the least square fits, all of them compatible with
the theoretical prediction $\tau \sim L$ to within $1.6\%$.}
\end{figure*}

To estimate the reflection and transmission coefficients within this
approximation, we require the solution of the diffusion equation with
absorbing boundary conditions at $x=0$ and $x=X$ and a constant unit
input flux at site $a$. In the steady state, the magnitude of the
fluxes at 0 and $X$ will give the splitting probabilities; i.e.  the
probability that a particle injected at position $a$ will be absorbed
at the origin or at $X$. In our scattering system the particles are
inciding on the left, which can be thought of as having the injection
point $a$ near the origin. Then the calculation outlined above yields
the transmission coefficient
\begin{equation}
T=a/X.
\label{trans}
\end{equation}

To estimate the mean survival time of the particles within the slab, we
recall that within the diffusion approximation, the survival time for
a random walker starting at position $a$ satisfies the equation~\cite{weiss}
\begin{equation}
D {d^2 \tau (a) \over d a^2}=-1~,
\end{equation}
with the conditions $\tau(0)=\tau(X)=0$, where $D$ is again the 
diffusion constant of the process. Thus
\begin{equation}
\tau(a)={1\over 2D}a(X-a)~,
\end{equation} 
and if the injection point is taken to be close to the origin 
($a\ll X$), we obtain 
\begin{equation}
\tau \sim aX/2D.
\label{time}
\end{equation}

These results can also be obtained on a more general footing through
Wald's identity ~\cite{weiss}, and are expected to hold as long as the
rms step size of the random walk is small compared to the
system size $X$, and the number of steps given by the random walker
scales linearly with time (i.e. there are no long tail waiting time
distributions).

We can identify the quantities appearing in Eqs.~(\ref{trans}) and
(\ref{time}) corresponding to the Lorentz scattering experiment.  The
diffusion coefficient $D$ is computed in Ref.~\cite{mac83} in a random
walk approximation and numerically through the Green-Kubo
relation. The length $X$ is related to $w$ and $L$ by
$X=L(2+w)\sqrt{3}/2$ and thus the penetration length $a$ can be
determined by the slope of the dependence of $\tau$ on $L$ as in
Fig.~\ref{ftau}. The results are consistent with the distance between
traps $(2+w)/\sqrt{3}$, as defined in Ref.~\cite{mac83}, for small
values of $w$ (where also the diffusion constant predicted in the
random walk approximation agrees with the numerically determined
one).

\section{Infinite horizon}
\label{infinite}

When the horizon becomes infinite the analogy to the simple diffusive
process breaks down. This occurs as a consequence of the opening of
infinite corridors between scatterers in which the particle is capable
of travelling very large distances between collisions.  The
distribution of the length of these sojourns, $p(r)$, has been shown
to decay as $r^{-3}$ when $r\to\infty$ by both numerical results and
theoretical arguments~\cite{bou85,zac86,ble92,dal97}.  Thus the rms
step length diverges and the diffusive approximation described in the
last section breaks down.

If we insist on making a random walk description of the system, we are
now led to consider a random walk with a distribution of step
lengths without second moment (generically called ``L\'evy flights''
~\cite{weiss,bou90}). Here one sometimes makes the distinction between
a discrete time process, in which each step is selected from a
power-law distribution of lengths, but always takes a fixed time and
the so-called L\'evy walk in which each step takes a time proportional
to its length. This latter, while more relevant to the case we are
discussing, is not significantly different from the L\'evy flight if
the first moment of the step size distribution exists, as it certainly
does in our case. We shall therefore ignore the considerable
complications this causes and often identify time with the number of
jumps or collisions $n$.

In contrast to the diffusive case, the derivation of the transmission
coefficient in the case of L\'evy flights appears not to have been
treated in the literature. We present an argument which leads to a
scaling prediction which seems reasonable for general L\'evy flights,
and specialize it to the case we are concerned with.

Denote the step distribution of the L\'evy flight as $P(r)\sim
1/|r|^{1+\alpha}$ with $0<\alpha\leq 2$.  A random walker with this
step distribution travels a typical distance \cite{bou90}
$$
\xi^2 \sim \cases {t^{2/\alpha} &{\rm for} $1<\alpha<2$\cr
                   t\ln t       &{\rm for} $\alpha=2$ \cr
                   t            &{\rm for} $\alpha>2$ \cr}
$$ 
in time t.

If we assume that the particle is initially at a point $a$
sufficiently near the origin ($a \ll X$), then we can invoke
Sparre-Andersen's theorem~\cite{fel}.  This theorem states that the
probability that the walker steps for the first time to the left of
the point $a$ at the $n$-th step is a universal function $\pi(n)$,
which is completely independent of any of the properties of $P(r)$, as
long as $P(r)$ is symmetric and continuous.  This distribution is
found to decay as $\pi(n)\sim n^{-3/2}$ for any kind of unbiased walk
whatsoever, in particular for the L\'evy flights we are concerned
with. This allows us to estimate straightforwardly the behaviour of
L\'evy flights both as regards their transmission and the mean
survival time in the interval considered.

The scaling behavior of the transmission coefficient of a L\'evy walk
across a finite system of length $X$ can be estimated as follows:
Denote by $\tau_\alpha(X)$ the time required for the walker to travel
a distance of order $X$ with appreciable probability. The transmission
coefficient is then the probability that the walker never steps left
of the origin during $\tau_\alpha(X)$. From the above scaling
relation, $\tau_\alpha(X)$ is expected to scale as
$X^\alpha$ for $1<\alpha<2$, as $X^2/\ln X$ for $\alpha=2$ and as
$X^2$ for $\alpha>2$, which corresponds to normal diffusion. Then, in
terms of $\tau_\alpha(X)$ the transmission coeficient is given by

\begin{equation}
T(X) \sim \sum_{n\geq \tau_\alpha(X)}^\infty\pi(n)
\sim \int_{\tau_\alpha(X)}^{\infty} t^{-3/2} dt 
\sim 1/\sqrt{\tau_\alpha(X)}.
\end{equation}

For $\alpha >2$ we therefore get the behavior of the ordinary
diffusive case treated in Section~\ref{wald}. In particular, for our
infinite horizon Lorentz slab, we expect $T(X)\sim {\sqrt{\ln L}}/L$.

As for the survival time, it is estimated in a similar way: The
probability of leaving the interval $[0,X]$ at time $n$ is given by
$\pi(n)$ as long as $n$ is not so large that leaving at the right
hand-side becomes appreciably likely.  This occurs at times of the
order of $\tau_\alpha(X)$.  This reasoning gives for the mean first
exit time $\tau$
\begin{equation}
\tau\sim\sum_{n=1}^{\tau_\alpha(X)}n\pi(n)\sim 
\sqrt{\tau_\alpha(X)}.
\end{equation}

\begin{figure*}
\vspace{6cm}
\includegraphics{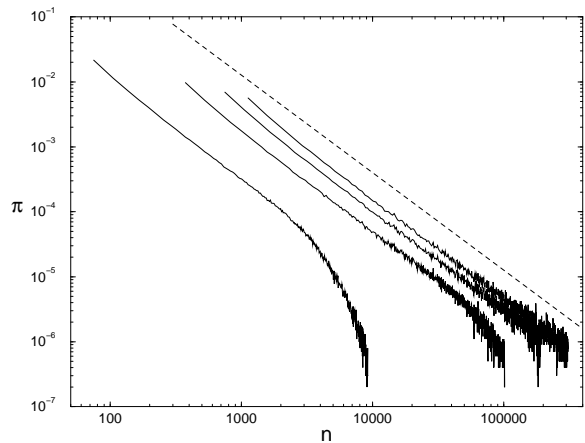}
\caption{ \label{pnc} The probability distribution $\pi (n)$
of leaving the system to the left after $n$ collisions, for $w=1.5$,
and from left to right $L=100, 500, 1000, 1500$ and $N=10^7$.  The
Sparre-Andersen scaling result is shown by the dashed lines.}
\end{figure*}

The prediction for the probability distribution $\pi (n)$ of leaving
the slab to the left after $n$ collisions based on the Sparre-Andersen
theorem can be tested for the Lorentz gas; it is valid both for the
finite horizon case and for the infinite horizon. In Fig.~\ref{pnc} we
show this distribution as obtained from the simulation 
for $w=1.5$ (infinite horizon) together with the
Sparre-Andersen scaling law. The agreement is good even for small
values of $n$.

\begin{figure*}
\vspace{6cm}
\includegraphics{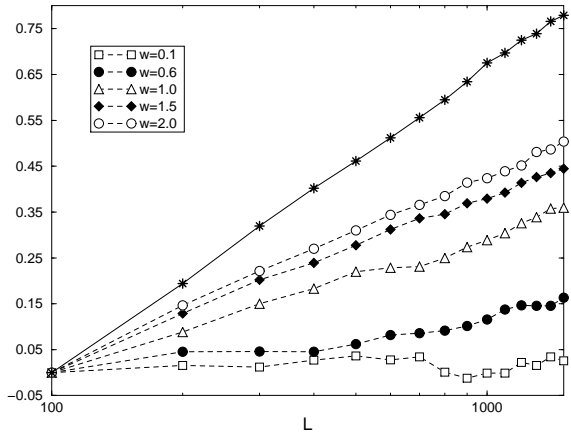}
\caption{ \label{ftl1} The logarithmic correction to $T$ is shown
by plotting
$[(LT(L))^2-(100T(100))^2]/(100T(100))^2$
vs. $\ln(L)$ for $w=0.1$ (with $N=4\times 10^7$), and $w=0.6, 1.0,
1.5, 2.0$ (with $N=10^7$). Straight lines with nonzero slope,
indicating the logarithmic correction, are expected for $w$ above the
infinite horizon. The dotted lines are meant as a guide to the eye.}
\end{figure*}

\begin{figure*}
\vspace{6cm}
\includegraphics{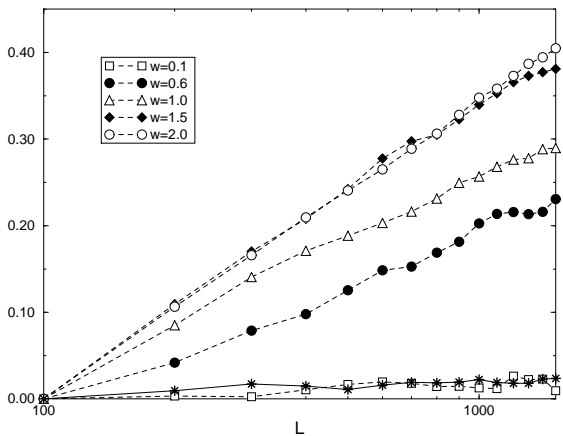}
\caption{ \label{ftau1}
The logarithmic correction to $\tau$ is shown by plotting
$[(L/\tau(L))^2)-(100/\tau(100))^2]/(100/\tau(100))^2$
vs $\ln(L)$ for the same
values of $w$ and $N$ as in Fig.~\protect{\ref{ftl1}}. Again straight
lines with nonzero slope are expected for $w$ above the infinite
horizon, and we also show the corresponding quantity por a L\'evy walk
with $\alpha=2.0$. The dotted lines are meant as a guide to the eye.}
\end{figure*}

The scaling laws for the transmission coefficient and for the mean
survival time can be also tested for the Lorentz gas in the infinite
horizon case.  In Fig.~\ref{ftl1} we show $(TL)^2$
for a set of $w$ values as a function of $\ln (L)$. According to our theory
$(TL)^2\sim\ln (L)$ for $w_c<w$. In Fig.\ref{ftau1} we show $(\tau/L)^2$
as a function of $\ln (L)$
in order put in evidence the
predicted logarithmic correction.  

Runs were also carried out at other fixed incidence angles, and the
same features as described above were obtained.  We have thus shown
that, given a fixed angle incidence, the opening of the horizon
appears to produce logarithmic corrections to the scaling laws present
for finite horizon. This feature is shared by the behavior of other
quantities which also present logarithmic corrections, such as the
diffusion coefficient.

\section{Angular Dependence of Forward and Backward Scattering}
\label{angular}

\begin{figure*}
\vspace{6cm}
\includegraphics{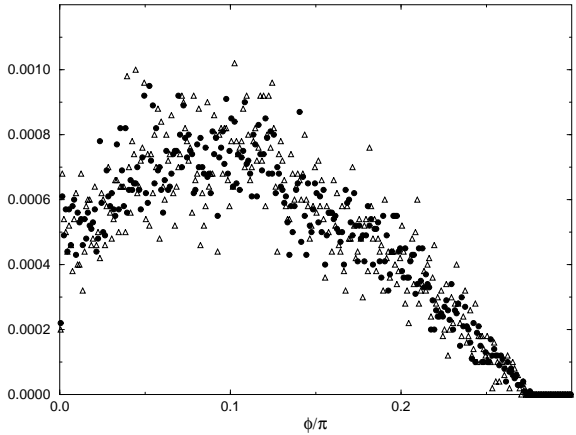}
\caption{ \label{symf} The quantities $b_T (\phi/\pi)$
({\Large $\bullet$}) and
$b_R ((\pi - \phi)/\pi)$ ($\bigtriangleup$) vs. $\phi/\pi$ for $w=0.3$ and
$N=10^7$. The values for $b_T$ are extracted from data with $L=100$
and those for $b_R$ are determined by substraction of data with
$L=200$ and $L=100$.}
\end{figure*}

\begin{figure*}
\vspace{6cm}
\includegraphics{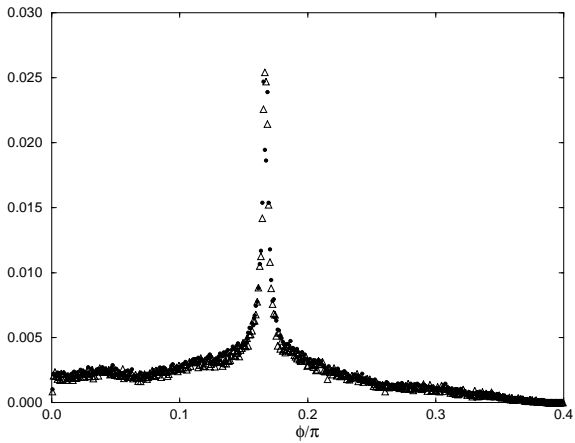}
\caption{ \label{symi}
The quantities $-b_T (\phi/\pi)$ ({\Large $\bullet$}) and
$b_R ((\pi - \phi)/\pi)$ ($\bigtriangleup$) vs. $\phi/\pi$ for $w=1.5$
and $N=10^7$. The values for $b_T$ are extracted from
data with $L=100$ and those for $b_R$ are determined
by substraction of data with $L=200$ and $L=100$.}
\end{figure*}

We have already observed that for fixed incidence angle the
transmission coefficient scales as $1/L$ in the finite horizon case,
whereas it scales as $\sqrt{\ln L}/L$ in the case of infinite
horizon. It is also instructive to look in somewhat greater detail at
the angular distribution of the transmitted and reflected
particles (this is what is done also in
chaotic scattering experiments with fewer scatterers~\cite{eck88}).
We define $\sigma_R(\phi)$ as the density of particles
reflected at angle $\phi$ and $\sigma_T(\phi)$ similarly for the
transmitted particles. Due to the symmetry $\phi \to - \phi$ the range
of $\phi$ will be from 0 to $\pi/2$ for transmitted particles and from
$\pi/2$ to $\pi$ for those that are reflected. As these distributions
are also dependent on $L$, we further propose that in the finite
horizon case they can be expanded as
\begin{equation}
\sigma_R(\phi)=a_R(\phi)+{b_R(\phi)\over L}+\cdots
\end{equation}
and 
\begin{equation}
\sigma_T(\phi)={b_T(\phi)\over L}+\cdots.
\end{equation}
The leading term in the transmission cross section is zero since there
is no transmission for infinite $L$. The following relation has then
been found to hold in all cases for sufficiently large $L$ (see
Fig.~\ref{symf}):
\begin{equation}
b_R(\phi)=-b_T(\pi-\phi)
\label{sym1}
\end{equation}
This symmetry relation can be understood as follows: Imagine that the
slab is subjected to a continuous flow of particles with identical
distributions incident from {\it both\/} sides of the
sample.  If we now look at the distribution of angles of the particles
which go to the right in these circumstances, it is given by the
superposition $\sigma_T(\phi)+\sigma_R(\phi)=
a_R(\phi)+{b_R(\phi)\over L}+{b_T(\phi)\over L}+\cdots$. Now, by
symmetry, the above set-up is equivalent to having particles incident
only from the right and a reflecting wall at the middle of the
sample.  However, from the fact that a particle travelling deep into
the slab eventually loses memory of its original incidence direction,
it follows that the trajectory of a particle reflected on the wall is
indistinguishable with the trajectory of a particle travelling in a
semi-infinite system and eventually returning across the position of
the wall (which occurs with probability one). Thus, given that the
correlation with the initial incidence is small enough, the system
with the reflecting wall will not show any great difference from the
semi-infinite system ($L=\infty$) as far as the angular distribution
of its particles is concerned. This then allows to derive the identity
\begin{equation}
a_R(\phi)=a_R(\phi)+{b_R(\phi)+b_T(\pi-\phi)\over L}+\cdots
\end{equation}
from which the result follows.  Numerically, this is quite well borne
out both in the case of finite (Fig.~\ref{symf}) and infinite horizon
(Fig.~\ref{symi}).

\begin{figure*}
\vspace{6cm}
\includegraphics{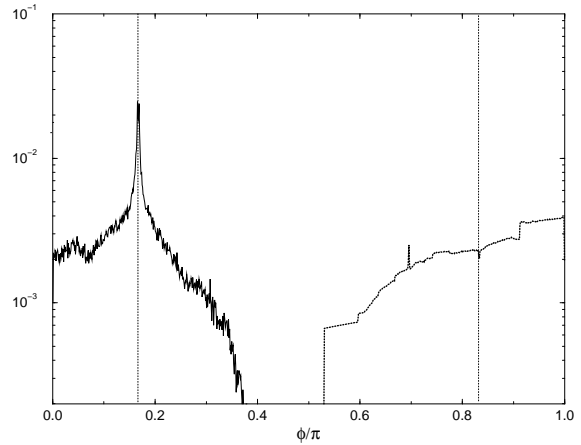}
\caption{ \label{dip} The quantities $b_T$ (full line) and
$a_R$ (dashed line) vs. $\phi/\pi$ for
$w=1.5$, $N=10^6$. The features expected at $\phi = \pi /6$ and
$\phi = 5\pi /6$ are clear. There is also a peak at
$\phi\approx {\rm arctan}\sqrt{2}$, for which we have no explanation.}
\end{figure*}

In the case of infinite horizon, another striking feature of the
angular distribution function is found: there is a clear peak in
$b_T$ around the value $\phi_c=\pi/6$ as well as a corresponding dip in
$a_R$ around the value $\phi_c=5\pi/6$ (see Fig.\ref{dip})
which are the angles the infinite corridors make with the edge of the
sample. Qualitatively, the appearence of these singularities can be
argued as follows. As is well-known, the periodic system corresponding
to the one we are studying is ergodic, so that all allowed positions
and all directions are eventually equally probable. In the finite
system we are studying, this is no longer the case.  In particular,
near the edges, the relative weight of the directions leading to
escape will be different from the other ones.  Nevertheless, we may at
first start with the approximation that, at least as long as the
particles enter reasonably deep into the system, the hypothesis of
equidistribution of velocity directions holds to a good
approximation.  Then we may estimate the transmission coefficient as a
function of angle by means of the fraction of surface area from which
trajectories escape at that angle (we may assume that the transverse
direction of the gas is made finite by some device such as periodic
boundary conditions).  This yields a singularity in the angular
dependence of the transmission coefficient near the critical angles
when $L\to\infty$.

\section{The Effect of Isotropic Incidence}
\label{isotropic}

The following apparent difficulty motivated us to also study the case
in which the particles are isotropically incident upon the slab: In
the preceding Section, we reported numerical and theoretical evidence
for the existence of a logarithmic correction to the mean survival
time. Yet a quite general argument appears to show such a correction
is impossible. Consider the phase space on a constant energy shell
inside the Lorentz array. The volume of this phase space, which
clearly scales as $L$, can be expressed as the integral of the time of
residence inside the system over all points of entry (this relation
is known as the Katz formula \cite{katz}). Since the volume of
trajectories that remain forever confined to the system is zero, and
the volumes of those entering from the left and from the right are
equal, it follows that the mean survival time scales as $L$, in
contradiction to what was numerically observed in the last Section, as
well as to the predictions of the L\'evy flight model.

If we consider the above argument carefully, however, we see that it
only applies if all angles of incidence are taken as equally
probable. To understand why this might make a difference, one should
note the following: If all angles are equally probable, then the
distribution of initial step lengths will have a singularity due to
the probability of launching the particle with an angle very close to
critical and an appropiate impact parameter. This leads, as is readily
seen, to the probability for a large initial step length of $x$ going
as $x^{-2}$. This behaviour is in marked contrast to the step
probability distribution within the interior of the system, for which
the impact parameter and the angle are interrelated. For these, as was
pointed out before, the probability of a large step $x$ goes as
$x^{-3}$.

In order to clarify these issues, we performed simulations on the system with
isotropic incidence. Indeed, as expected from the above argument,
the mean survival time was found to scale as $L$, without any
logarithmic corrections (see Fig.~\ref{ftau1}). 
On the other hand, the transmission
coefficient was found to retain its peculiar 
behaviour (see Fig.~\ref{ftl1}).  The above
argument is therefore sound, it does not contradict the
numerical work reported in the previous Section. On the contrary,
since the two models show such clear differences with regard to their
mean survival time, this seems to indicate that the distribution of
initial step lengths may well have been the cause.

To test this final hypothesis, we also simulated a L\'evy walk with
$\alpha=2$ in which the first step has a distribution with an $x^{-2}$
tail, corresponding to $\alpha=1$.  The results are again very clear:
the mean survival time now grows as $L$ without any logarithmic
corrections.  On the other hand, the transmission coefficient still
has the previous anomalous behaviour, though it takes a somewhat
longer time to reach it.

\section{Concluding remarks}
\label{conclusions}

In this work we have examined the scattering properties of a Lorentz
gas incident on an array of scatterers centered on a triangular
lattice with a finite number $L$ of columns. Much of our attention has
focused on the scaling with $L$, for large values of $L$, of transport
and optical properties, namely, transmission and reflection
coeficients, mean survival time and differential cross section.  It
should be emphasized that though most of the numerical results
reported in this paper were obtained for incidence along the $x$
direction, exactly the same phenomena was observed in runs carried out
at other fixed incident angles. On the other hand, some significant
differences were found when the particles were launched isotropically
in the infinite horizon case.

For our understanding of the observed numerical trends we have
considerably profited from the link, in some cases formal, in others
qualitative, to random walk processes.  Two regimes are considered:
the finite and the infinite horizon cases. For finite horizon the
transmission coefficient scales with $1/L$, the survival time with $L$
and the differential cross-section has no singularities and presents
certain symmetry properties.  All this is in agreement with the
behavior of normal diffusion and ordinary random walks with absorbing
boundaries.

The infinite horizon case, on the other hand, exhibits logarithmic
corrections to the aforementioned scalings depending on the incidence
angle distribution (in this case the relation to L\'evy walks is
illuminating). Also, singularities appear in the differential cross
section at angles corresponding to the corridors inside the slab, for
which we only have a partial understanding.  A peculiar effect related
to the difference observed between particles launched in one fixed
direction and particles launched isotropically can also be explained
in terms of a simple L\'evy flight model with a different distribution
for the initial step length.  This allows an explanation for the a
discrepancy in the behavior of the mean survival time, which had been
anticipated on quite general grounds.

It is also of interest to understand the features of the free
motion length distribution. For finite horizon two peaks in such
distribution are present at the values $w$ and $\sqrt{3} (2+w) -2$,
which correspond to the unstable periodic orbits perpendicular to the
disks.  For infinite horizon a set of peaks develops, whose number
increases with $w$; the slope of the envelope of the probability
distribution is $-3$. The origin of such peaks, which most probably
are related to other periodic orbits, remains to be analyzed.

\section*{Acknowledgements}

We thank R. Artuso, L. Benet, P. Dalquist, J. L. Lebowitz, C. Mej\'{\i}a,
T. H. Seligman, and L. A. Torres for
fruitful discussions and suggestions. This work has been partially
supported by INFN, CONACyT, DGAPA-UNAM under contracts IN103595, IN106597
and CIC, Cuernavaca. It is also part of the European Contract
No. ERBCHRXCT940460 on ``Stability and universality in classical
mechanics''.


\end{document}